# Tailoring Exciton Dynamics in TMDC Heterobilayers in the Quantum Plasmonic Regime

Mahfujur Rahaman,[1] Gwangwoo Kim,[1] Kyung Yeol Ma,[2] Seunguk Song,[1] Hyeon Suk Shin,[2] and Deep Jariwala[1]*

[1]Department of Electrical and Systems Engineering, University of Pennsylvania, Philadelphia, PA 19104, USA

[2]Department of Chemistry, Ulsan National Institute of Science and Technology (UNIST), UNIST-gil 50, Ulsan 44919, Republic of Korea

*Corresponding author: dmj@seas.upenn.edu (D.J.)

**Abstract**

Control of excitons in transition metal dichalcogenides (TMDCs) and their heterostructures is fundamentally interesting for tailoring light-matter interactions and exploring their potential applications in high-efficiency optoelectronic and nonlinear photonic devices. While both intra- and interlayer excitons in TMDCs have been heavily studied, their behavior in the quantum tunneling regime, in which the TMDC or its heterostructure is optically excited and concurrently serves as a tunnel junction barrier, remains unexplored. Here, using the degree of freedom of a metallic probe in an atomic force microscope, we investigated both intralayer and interlayer excitons dynamics in TMDC heterobilayers via locally controlled junction current in a finely tuned sub-nanometer tip-sample cavity. Our tip-enhanced photoluminescence measurements reveal a significantly different exciton-quantum plasmon coupling for intralayer and interlayer excitons due to different orientation of the dipoles of the respective *e-h* pairs. Using a steady-state rate equation fit, we extracted field gradients, radiative and nonradiative relaxation rates for excitons in the quantum tunneling regime with and without junction current. Our results show that tip-induced radiative (nonradiative) relaxation of intralayer (interlayer) excitons becomes dominant in the quantum tunneling regime due to the Purcell effect. These findings have important implications for near-field probing of excitonic materials in the strong-coupling regime.

**Introduction**

Coulomb bound electron-hole (*e-h*) pairs, commonly known as excitons, govern the optical properties of monolayer transition metal dichalcogenides (TMDCs) due to their large binding energies (on the scale of 0.5 eV) and oscillator strengths[1]. As a result, the fundamental optical properties of these materials are dominated by many body excitonic resonances, even at room temperature (RT). Furthermore, in a homo/hetero-bilayer (HBs) sample made from TMDCs, ultrafast interlayer charge transfer can also facilitate the formation of interlayer excitons (ILXs) with long lifetimes and large exciton binding energies observed at RT in prior work[2]. Therefore, TMDCs have attracted significant attention for both fundamental studies of novel quantum optical phenomena and photonic/optoelectronic applications in recent times[3–9].

TMDCs possess strong light-matter coupling at excitonic resonances in the visible part of the spectrum, with almost ideal two-dimensional (2D) confinement, making it easier to control the excitonic parameters such as resonance energies, oscillator strength, radiative and nonradiative lifetimes on demand[10–14]. Therefore, one way of controlling excitons is to manipulate them via plasmonic coupling, using plasmonic resonances in noble metal nanostructures which are also in the visible spectrum. In particular, the use of metallic nanostructures in the proximity of TMDC monolayers can create both weak and strong coupling regimes for excitons, and thus, control the emission energies, decay rates, radiative, and nonradiative lifetimes[15–17].

In general, excitons in TMDCs in proximity to a plasmonic system can be treated as dipole emitters, whose emission can be expanded into multipoles centered around the plasmonic energy[18]. Hence, the strength of coupling between the exciton and plasmon, and the associated manipulation of excitonic parameters, depends on the individual field polarizability. Therefore, the manipulation of intralayer (in-plane polarization) and interlayer (out-of-plane polarization) excitonic parameters in TMDC HBs via a plasmonic cavity can be different due to their different polarization states. ILXs, in particular, show great tunability in a plasmonic cavity as a function of cavity size in the $z$-direction (*i.e.*, coupling efficiently to the ILX polarization), resulting in the amplification of both exciton decay rate and radiative lifetime[19,20]. However, as the cavity size is further tuned from nanometer to sub-nm gap (in the quantum plasmonic regime), a strong interaction between the plasmonic field and ILX results in more nonradiative loss. Whereas, in-plane polarized intralayer excitons show an opposite trend as the size of the cavity decreases further in the sub-nm scale due to the Purcell effect[21].

The ultrafast exciton-plasmon interaction dynamics are generally probed via pump-probe and time-resolved PL measurements in a conventional optical configuration in the form of overall PL lifetimes of the excitonic species[22,23]. However, it is not feasible to finely control cavity size in the sub-nm scale and simultaneously probe the exciton-plasmon interaction dynamics, let alone discern radiative and nonradiative contributions using a conventional setup. Recently, a qualitative approach of determining the individual contribution of radiative and nonradiative decays and the Purcell effect on intralayer and interlayer excitons in TMDC HBs has been proposed using tip-enhanced photoluminescence (TEPL) configuration in a finely tuned sub-nm cavity[21]. In this approach, the PL measured from both intralayer and interlayer excitonic emissions in a sub-nm cavity can be fitted using a rate equation model to decern the contribution of Purcell enhancing/quenching, radiative, and nonradiative lifetimes. Although the model, effectively deconvoluted all the contributing parameters, an important question remained unanswered, which is how these parameters evolved in a quantum plasmonic regime (sub-nm cavity) when the junction current flows through the channel. This is particularly relevant since previous works have predicted that tip-induced tunneling through the TMDC monolayers to the metal substrate can reduce the strength of the plasmonic field in the sub-nm cavity and hence decrease the intralayer excitonic emission[24,25].

Here, we conduct a systematic investigation into the effect of junction current on the dynamics of intralayer and interlayer exciton-plasmon interactions in TMDC HBs within the quantum plasmonic regime, using a finely tune tip-sample cavity in a TEPL configuration. We utilize $MoS_2$/$WSe_2$ HBs as a test bench on a hBN/Au substrate. Our findings indicate that as the tip-sample distance decreases below 1 nm, in the absence of junction current, intralayer exciton amplifies while ILXs decrease drastically due to the Purcell effect and stronger nonradiative coupling to the plasmon field respectively. Moreover, once a channel is established for current to flow through the HB/hBN to the Au substrate, a reverse trend is observed. Using a rate equation model, we qualitatively determined all the coupling parameters, including the dynamics of exciton-plasmon interactions. To the best of our knowledge, our results present the first experimental demonstration of the dynamics of exciton-plasmon interactions in the presence of junction current in the quantum plasmonic regime.

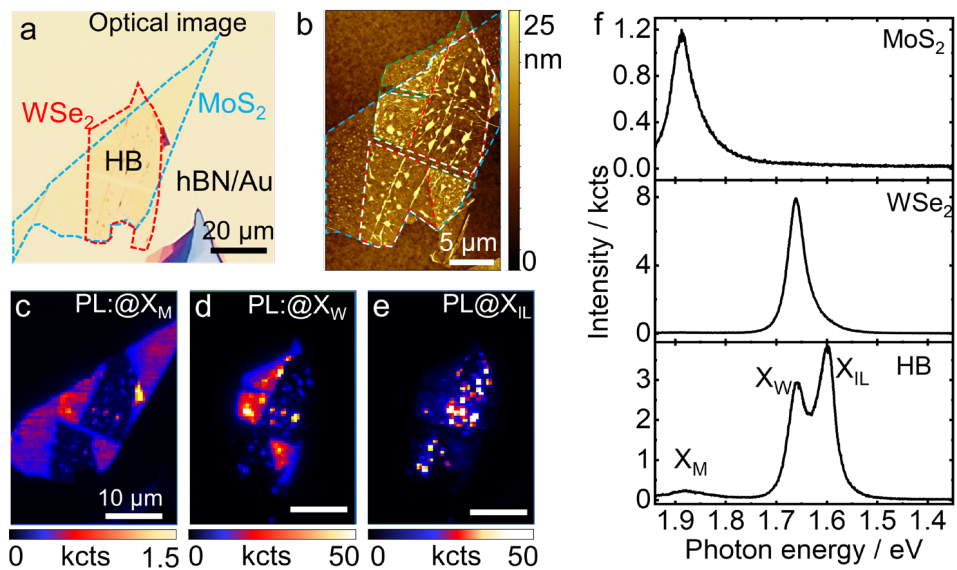

**Figure 1: Far-field optical characterization of HB.** (a) Optical image and (b) AFM topography of one of the MoS$_2$/WSe$_2$ HB sample prepared on 3 nm hBN/Au substrate. (c) – (e) PL maps of intralayer exciton MoS$_2$ (X$_M$), WSe$_2$ (X$_W$) and interlayer exciton (ILX) across MoS$_2$/WSe$_2$ interface respectively. Areas where strong interfacial coupling is established ILX have strong emission followed by quenching of intralayer X$_M$ and X$_W$. (f) Three representative PL spectra of MoS$_2$, WSe$_2$, and HB regions displaying characteristic PL spectra of intralayer and interlayer excitons.

## Results and Discussions

Fig. 1a,b show an optical image and atomic force microscope (AFM) topography, respectively, of one of the MoS$_2$/WSe$_2$ HB samples prepared on hBN/Au substrate. Details of the sample preparation can be found in the experimental section and supplementary information (S1). Three far-field PL maps are created: two for intralayer excitons X$_M$ (monolayer MoS$_2$) and X$_W$ (monolayer WSe$_2$), and one for interlayer exciton X$_{IL}$ (HB) across the MoS$_2$/WSe$_2$ interface, respectively, and presented in Fig. 1c-e. Three corresponding far-field PL spectra are displayed in Fig. 1f. As shown in the AFM topography and PL maps, areas marked by the red dashed lines only produce strong ILXs, suggesting better interfacial coupling. We also perform complementary surface potential mapping with/without illumination to further validate our hypothesis. Details of the Kelvin probe force microscope (KPFM) measurements can be found in the supplementary information (S2). We observe a strong ILX emission followed by heavily quenched intralayer X$_M$ and X$_W$ emission on the areas marked by red dashed lines, a hallmark of the ILX formation process.

After initial far-field characterization, we perform TEPL measurements on areas of strong interfacial coupling. Fig. 2a shows a schematic of the TEPL measurements. We use an Au tip for the TEPL measurements under 633 nm excitation. The introduction of an Au substrate creates a plasmonic dimer cavity, the polarization of which is perpendicular to the basal plane of the HB (as shown by ***E*** in the scheme). We also tune the tip-sample distance (*d*) via AFM piezo actuator to investigate exciton dynamics in the HBs. We use 3 nm hBN grown by chemical vapor deposition (CVD) as the insulating barrier between HBs and the substrate. Fig. 2b displays an AFM topography image taken across the boundary of HB and WSe$_2$. The white dashed line is drawn as a guide to the eye along the border line. A TEPL hyperspectral map is acquired across the boundary and superimposed on the corresponding topography area within the AFM image in Fig. 2b. The TEPL map is created for $X_{IL}$ spectral range. Two representative TEPL spectra of the two regions (red and blue circles on the TEPL map) are shown in Fig. 2c. The orange rectangular shade is the spectral area for which the $X_{IL}$ map is created in Fig. 2b. As can be seen, TEPL spectra of HB is dominated by ILX emission, with both intralayer $X_M$ and $X_W$ strongly quenched. Additionally, the TEPL map also exhibits a spatially homogeneous distribution of the $X_{IL}$ intensity in the HB region.

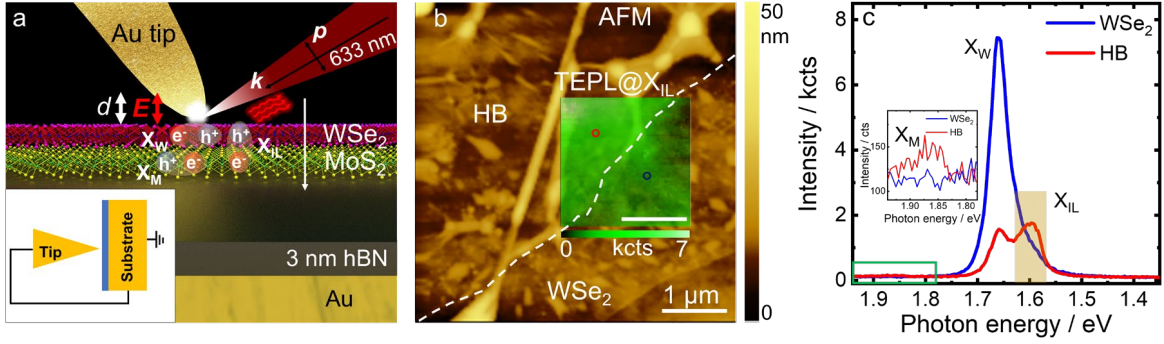

**Figure 2: TEPL study of HB.** (a) Schematic illustration of TEPL measurements. In-plane intralayer $X_M$ and $X_W$ and out-of-plane interlayer $X_{IL}$ were excited/amplified by the plasmonic field created at the tip apex under 633 nm excitation. Introduction of a Au substrate created a dimer cavity with the polarization direction perpendicular to the basal plane of the HB. Tip-sample distance was tuned via AFM piezo actuator from few nm to sub-nm gap and TEPL signal was collected. Inset: the schematic of the electrical configuration of the tip-sample junction. (b) AFM topography image at the boundary of the HB and WSe$_2$. Inset: a TEPL map acquired for $X_{IL}$ across the boundary superimposed on the corresponding topography area. (c) Two representative TEPL spectra on the map taken from red and blue circles marked on the TEPL map image. Orange shade on the TEPL spectra is the spectral region for which the TEPL map was created. Inset: zoomed in spectral range covered by the green box highlighting $X_M$.

In order to investigate sub-nm tip-sample gap dynamics of exciton-plasmon interaction for both intra- and interlayer excitons in HB, we acquire TEPL spectra as a function of tip-sample distance at each point. In addition, we simultaneously record the junction current profile (current flowing from the tip to the substrate through the HB) as a function of tip-sample distances. The current profile is recorded in the short circuit configuration (*i.e.* tip and substrate are electrically connected and the bias voltage, V = 0 V as shown in the inset of Fig. 2a). Therefore, the driving force for the current flow in the sub-nm gap (quantum plasmonic regime) can be a combination of the tunneling of tip hot electrons through HB to the Au substrate and the photovoltage created at the HB interface under 633 nm excitation[26,27]. Important to note that, we consistently observe junction current at random points on the HB/hBN/Au sample. As mentioned earlier, we use a CVD-grown 3 nm thick hBN film as the insulating barrier between the HB and Au substrate. During the transfer process of the CVD-grown hBN film onto the Au substrate using the PMMA-assisted wet transfer method from the sapphire substrate (see experimental section), it is possible that the quality of the film is compromised, and random channels are opened for the current flow between the tip and substrate. To support our hypothesis, we also perform conductive AFM mapping on hBN/Au areas adjacent to the HB. Results of the conductive AFM mapping of hBN film are presented in the supplementary information (S3).

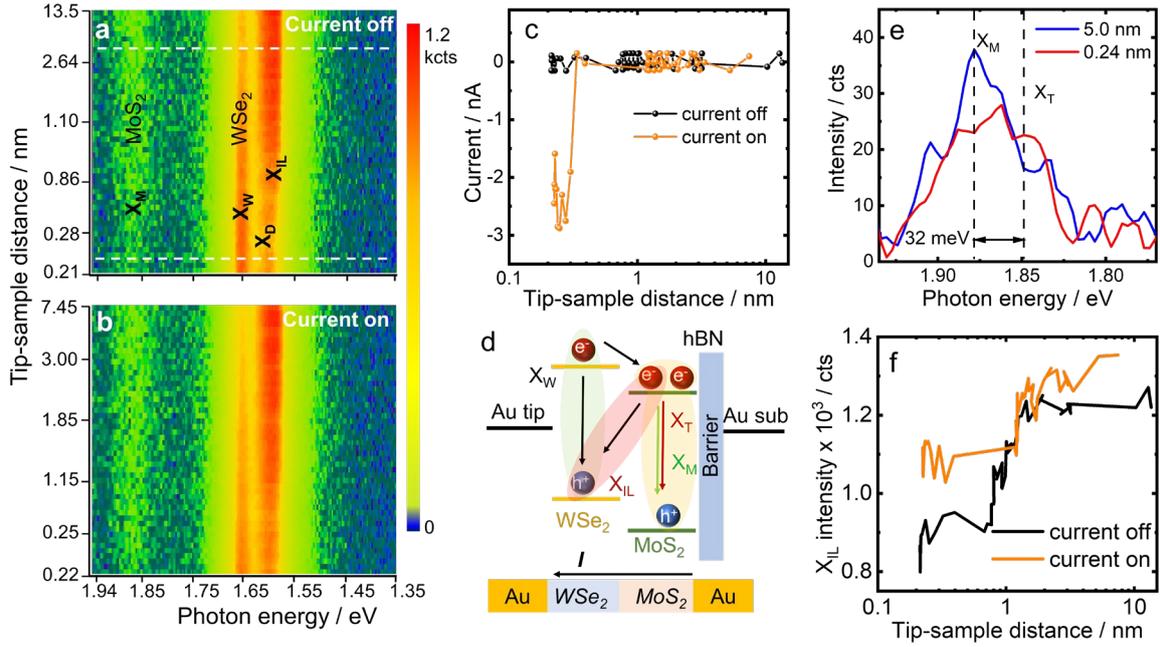

**Figure 3: Exciton tuning in the quantum tunnelling regime.** Spectral evolution of TEPL signal as a function of tip-sample distance (a) when no current flows through the HB and (b) when current flows through the HB. (c) Junction current profile as a function of tip-sample distance for the case of (a) and (b). Electrical configuration of the tip-sample junction is shown in Fig. 2a. Current was measured simultaneously in the short circuit configuration (V = 0 V). (d) Schematic of the HB band alignment showing trion formation in MoS2 and direction of current flow when the tip is in the sub-nm gap. (e) Comparison of MoS2 TEPL spectra at two different tip-sample distances (white dashed lines in (a)) for the case of no junction current. In addition to the $X_M$, we could also observe trion, $X_T$ in MoS$_2$. (f) $X_{IL}$ evolution as a function of tip-sample distance for the two cases (current off and on).

Fig. 3a,b show two sets of TEPL evolution as a function of tip-sample distance with junction current off and on respectively. These two data sets are recorded on two different points in the same TEPL map shown in Fig. 2b. The corresponding current vs tip-sample distance graphs are presented in Fig. 3c. Since electrons are flowing from the tip to the substrate (as shown in a schematic in Fig. 3d), and the substrate is grounded, we observe a negative current as a function of tip-sample distance. For the tip-sample distance-dependent study, we vary the AFM piezo actuator and record the corresponding Force curves, from which the actual tip-sample distances are calculated. Details of the tip-sample distance determination procedure can be found in the supplementary information (S4). PL evolution without junction current reveals two distinct tip-sample gap regimes: (i) in the nm gap (> 1 nm) regime all exciton intensities are increasing, and (ii) in the sub-nm gap regime intralayer (interlayer) exciton intensity is increasing (decreasing). Moreover, we can also observe that in the sub-nm gap $X_M$ intensity gradually deceases with gap size. Additionally, the contribution from dark exciton ($X_D$) of WSe$_2$ becomes apparent as the gap shrinks. Two representative TEPL spectra one in the nm gap and the other in the sub-nm gap regime are plotted in the supplementary information (S5) for the PL evolution map shown in Fig-3a. Observation of dark excitons in WSe$_2$ in the TEPL configuration is a well-known phenomenon, which originates from the radiative exchange between the exciton dipole and the tip plasmon[28,29]. However, the decreasing trend of $X_M$ may lie in the exciton population and interfacial charge transfer process, as the tip-sample gap shrinks. A schematic illustration of the exciton population

and relaxation process in the HB in the tip-sample gap is shown in Fig. 3d. Excitons are populated in both monolayers by gap plasmon excitation. Ultrafast interfacial charge transfer allows electrons in WSe$_2$ to cross the interface and jump to the conduction band of MoS$_2$. Since hBN acts as the barrier for electrons to move to the Au substrate, overall electron concentration may increase momentarily in MoS$_2$. This results in the radiative relaxation of ILX across the interface and formation of trion in MoS$_2$. Hence, we observe a gradual decrease in X$_M$ intensity as the gap shrinks. Fig. 3e displays two TEPL spectra in the MoS$_2$ spectral regime taken along the white dashed lines in Fig. 3a. As it is seen, PL spectra at 0.23 nm tip-sample distance clearly shows an overall broad spectrum with a trion peak at 35 meV[1] below the main excitonic peak in MoS$_2$, and supports our hypothesis.

An interesting phenomenon is observed when the junction current flowed (Fig. 3b) between the tip and the sample, especially in the sub-nm gap (quantum plasmonic regime). Both the intralayer exciton Purcell enhancement and ILX showing a decreasing trend are slowed down as the current started flowing in the shrinking gap. The evolution of ILX intensity as a function of gap size for both current off and on is shown in Fig. 3f for comparison. PL enhancement in the tip-sample cavity (tip-sample gap plus the HB/hBN thickness) involves a competition between the Purcell effect and the tip-induced nonradiative quenching[18]. Both the Purcell enhancement and the tip-induced nonradiative damping are scaled to a power law of the cavity size. Especially, nonradiative relaxation becomes significant in the sub-nm tip-sample gap via dipole coupling to the tip-sample cavity plasmon due to the ultrafast ohmic Drude loss[30]. Therefore, in the sub-nm gap, we observed a sharp rise of X$_W$ emission due to the Purcell effect and ILX quenching since tip-induced nonradiative damping of intralayer excitons becomes faster than the interlayer charge transfer. However, as soon as the current starts flowing between the tip and the sample, the strength of the cavity plasmon weakens. This results in reduced Purcell enhancement and slower nonradiative damping of intralayer excitons resulting in the boosting of ILX emission in the sub-nm gap. To the best of our knowledge, our results are the first experimental demonstration of exciton-plasmon coupling in the presence of junction current recorded in a DC-biased near-field spectroscopy experiment.

To understand the tip-sample gap induced contribution of radiative and nonradiative damping as well as the near-field enhancement, we fit the PL evolutions using a steady state rate equation model described in the previous work[21]. Details of our model and fitting procedure are discussed in the supplementary information (S6). Evolution of all excitonic populations is the product of various excitation and relaxation rates inside the cavity. The cavity induced field enhancement (excitation) can be scaled as $F \propto (R/z)^m$, with $R$ being the tip radius, $z$ being the distance between the tip and Au substrate, and $m$ is the geometrical factor. In contrast, population of ILX depends on the interlayer charge transfer upon intralayer exciton population. We divide the model into two regions: one for the nm gap and the other for the sub-nm gap, with only adjusting parameter is the scaling factor. The total decay rate of each excitonic species is the sum of three terms: (i) cavity-controlled (Purcell effect) radiative decay scaled as $\Gamma^{rad} \propto (z+d)^{-n} + \Gamma_0^{rad}$, with $d$ is the minimum tip-sample distance, $n$ is the scaling factor, and $\Gamma_0^{rad}$ is the free space radiative decay; (ii) cavity-induced nonradiative recombination described by $\Gamma^{nrad} \propto (R/(z+d))^l$, with $l$ being the scaling factor; and (iii) first-order intrinsic nonradiative relaxation rate $\Gamma_0^{nrad}$. The value of $\Gamma_0 = 2\hbar/\tau_0$, with $\tau_0^{rad}$ and $\tau_0^{nrad}$ are assumed to be 0.7 ns and 1.5 ps respectively (taken from ref[21,31]). Using these assumptions in the steady-state limit of exciton population, we fit the PL evolution of X$_W$ for both sets of results shown in Fig. 3a,b. We are not able to extract the X$_M$ intensity profile reliably due to its very low quantum yield. Therefore, we do not fit tip-substrate cavity dependent X$_M$ evolution in the present study. Moreover, since ILX population requires contribution from both $X_M$ and $X_W$, we need fit parameters from both intralayer excitonic species for a reliable fitting. Hence, we also avoid any qualitative analytical discussion on ILX parameters as well. Fig. 4a,b present fitted $X_W$ evolution as a function of tip-sample distance for the PL evolution graphs shown in Fig. 3a,b respectively. The PL evolution reveals two distinct distance-dependent regimes, which our model fits well. Since the model requires analytical expression of both radiative and nonradiative relaxation rates, it is possible to extract radiative and nonradiative lifetimes of fitted excitons in the varying tip-sample cavity. A qualitative discussion on radiative and nonradiative relaxation of $X_W$ in the tip-sample cavity is presented in the supplementary information (S6). Here, the evolution of the Purcell factor in the tip-substrate cavity is going to be discussed. Fig. 4c shows the evolution of the Purcell factor in the tip-substrate cavity extracted from the fitting presented in Fig. 4a,b respectively. Our model provided a similar scaling exponent to the model described in ref.[21] for the Purcell enhancement in the absence of junction current (see Table I in the supplementary information). However, a more dramatic change can be seen in the case of the current on. The cavity-dependent field enhancement initially increases at a scaling exponent of 5.6. However, as soon as the current starts flowing field strength is suppressed by an exponential factor of 0.5. The maximum Purcell factor is extracted to be $F \approx 6 \times 10^3$ calculated for the case of current off is consistent with previous TEPL measurements[32–34]. Additionally, the exponent factor $m \approx 5$ indicates that our near-field geometry is more like a point dipole on a plane for which the Purcell factor is expected to grow as $1/z^6$ [35]. This is most likely due to the fact that 3 nm hBN film on top of Au substrate results in reduced coupling between the tip and the Au substrate.

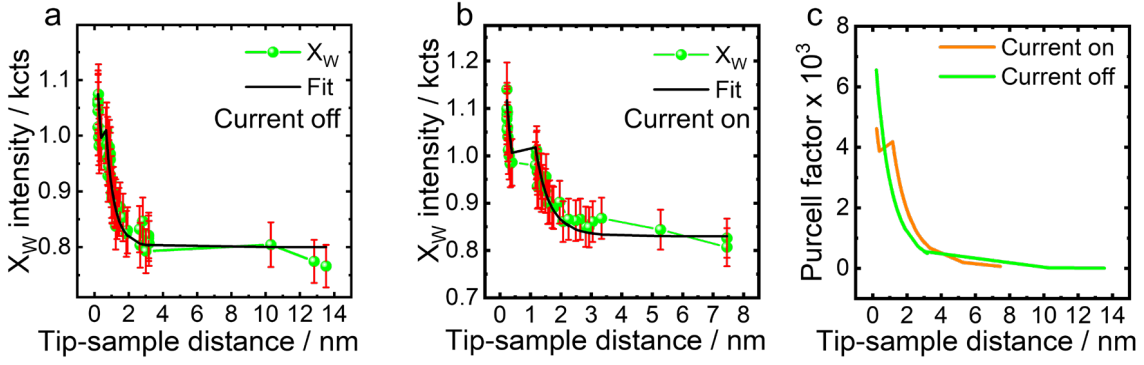

**Figure 4: Rate equation fit to the PL evolution.** PL evolution of intralayer $X_W$ together with fitted curve as a function of tip-sample cavity for (a) no current in the junction and (b) current flowing in the junction. (c) Calculated Purcell factor in the tip-sample cavity with/without junction current.

## Conclusion

In summary, this work reports on tailoring exciton dynamics in the near-field from the classical plasmonic regime (few nm) to quantum plasmonic regime (sub-nm) with/without junction current in TMDC HBs using an Au tip + Au substrate-induced plasmonic cavity. We show that in the absence of a junction current intralayer and interlayer exciton show an opposite trend as a function of gap size in the sub-nm cavity. We explain this behavior by two competing phenomena. While the cavity field amplifies intralayer excitons dramatically in the quantum plasmonic regime, it also enhances nonradiative damping via coupling between exciton dipole and tip plasmon for which interlayer exciton suffers the most. In contrast, when current flows in the junction, it quenches the Purcell factor of the sub-nm cavity dramatically, and at the same time boosts ILX by reducing the nonradiative relaxation of excitons. Our work provides a solid understanding of exciton dynamics in the quantum plasmonic regime with/without junction current and demonstrates a clear pathway of boosting exciton densities to enable new optoelectronic applications, and to induced room temperature exciton condensates via tuning plasmonic cavity in the quantum tunneling regime.

**Experimental Section**

$MoS_2$/$WSe_2$ HBs were prepared using PDMS assisted deterministic dry transfer method. Since interface contamination is one of the major challenges for the ILXs formation, we used PDMS to PDMS pick up/creation of HBs. Details of the HBs creation are schematically presented in the supplementary information section, S1. HBs prepared in this way show strong ILX emission as shown in Fig.1. We used CVD-grown 3 nm thick hBN film on top of a 100 nm thick Au film as the substrate. The 3 nm thick hBN film was prepared by a low-pressure CVD system on a c-plane sapphire substrate using ammonia borane as a precursor. The details of the CVD procedure of hBN and PMMA-assisted wet transfer of hBN on arbitrary substrates can be found in the literature[36]. After the preparation of the individual HBs, the bilayer stacks were then transferred onto hBN/Au substrate.

Far-field optical measurements were conducted using a Horiba LabRam HR evolution confocal microscope coupled with an electron multiplying charged couple detector dispersed by a 100 l/gr grating. A 633 nm solid-state laser was used for the excitation with laser power of 17 µW focused onto the sample surface via 100x 0.9 NA objective.

TEPL measurements were performed using a Horiba NanoRaman platform in the side illumination/collection configurations, which consists of an atomic force microscope from AIST-NT and a LabRam Evolution spectrometer. The Au tips used in the experiments were purchased from Horiba and suited for near-field measurements under 633 nm excitation. The laser power was kept at 17 µW and focused onto the tip via a 100x 0.7 NA long working distance objective. The exposure time was 0.2 s. TEPL hyperspectral maps were acquired in the spectop mode (a contact/noncontact hybrid mode developed by Horiba), in which half of the time (t1) the tip is in contact with the sample to acquire the near-field signal and the rest half of the time (t2) the tip is operating in the intermittent contact mode to acquire the far-field signal and the AFM topography on each pixel with the total time defined by t = t1+t2 and t1=t2=exposure time.

For tip-sample distance-dependent TEPL measurements, a varying DC voltage was applied gradually to the piezo-actuator connected to the sample stage for fine-tuning of the tip-sample cavity from few nm to sub-nm. A total of 50 data points were collected for a large piezo-actuator displacement (150 nm – 200 nm). At each point, a TEPL spectra was acquired using an exposure time of 0.2 s. In addition to the TEPL spectra, force vs distance curves and junction current were also monitored for the mentioned piezo-actuator tuning range for each experimental data set. The actual tip-sample distance, d was calculated from the acquired force vs distance curve. Details of the calculation can be found in the supplementary section, S2.

KPFM measurements were performed using AFM from AIST-NT and commercially available Cr/Au probes with/without illumination under 633 nm excitation.

## Competing Interests

The Authors declare no Competing Financial or Non-Financial Interests.

## Data Availability

The data that support the findings of this study are available on request from the corresponding author.

## Authors Contributions

M.R. and D.J. conceived the idea and designed the research. M.R. implemented the project via performing the experiments and simulations with the help of G.K and S.S. G.K. assisted in sample preparation. K.Y.M and H.S.S. prepared the hBN film. M.R. and D.J. wrote the manuscript; all the authors revised and commented on the manuscript. All authors contributed to the writing of manuscript and interpretation of the data.

Corresponding author: Deep Jariwala: dmj@seas.upenn.edu

## Acknowledgement

D.J. acknowledges primary support for this work by the Air Force Office of the Scientific Research (AFOSR) FA2386-20-1-4074. M.R. acknowledges support from Deutsche Forschungsgemeinschaft (DFG, German Research Foundation) for Walter Benjamin Fellowship (award no. RA 3646/1-1). G. Kim acknowledge primary support for this work by the Asian Office of Aerospace Research and Development (AOARD) of the Air Force Office of Scientific Research (AFOSR) FA2386-20-1-4074. The sample fabrication, assembly and characterization were carried out at the Singh Center for Nanotechnology at the University of Pennsylvania which is supported by the National Science Foundation (NSF) National Nanotechnology Coordinated Infrastructure Program grant NNCI-1542153. K.Y.M. and H.S.S. acknowledge the support from the National Research Foundation, Republic of Korea via the research fund (NRF-2021R1A3B1077184). S.S. acknowledges support from Basic Science Research Program through the National Research Foundation of Korea (NRF) funded by the Ministry of Education (Grant number 2021R1A6A3A14038492).

## References

(1) Kylänpää, I.; Komsa, H. P. Binding Energies of Exciton Complexes in Transition Metal Dichalcogenide Monolayers and Effect of Dielectric Environment. *Phys. Rev. B - Condens. Matter Mater. Phys.* **2015**, *92*, 1–6.

(2) Jin, C.; Ma, E. Y.; Karni, O.; Regan, E. C.; Wang, F.; Heinz, T. F. Ultrafast Dynamics in van Der Waals Heterostructures. *Nat. Nanotechnol.* **2018**, *13*, 994–1003.

(3) Liu, Y.; Weiss, N. O.; Duan, X. X.; Cheng, H. C.; Huang, Y.; Duan, X. X. Van Der Waals Heterostructures and Devices. *Nat. Rev. Mater.* **2016**, *1*.

(4) Ross, J. S.; Rivera, P.; Schaibley, J.; Lee-Wong, E.; Yu, H.; Taniguchi, T.; Watanabe, K.;

Yan, J.; Mandrus, D.; Cobden, D.; Yao, W.; Xu, X. Interlayer Exciton Optoelectronics in a 2D Heterostructure P-n Junction. *Nano Lett.* **2017**, *17*, 638–643.

(5) Combescot, M.; Combescot, R.; Dubin, F. Bose-Einstein Condensation and Indirect Excitons: A Review. *Reports Prog. Phys.* **2017**, *80*.

(6) Jiang, C.; Xu, W.; Rasmita, A.; Huang, Z.; Li, K.; Xiong, Q.; Gao, W. B. Microsecond Dark-Exciton Valley Polarization Memory in Two-Dimensional Heterostructures. *Nat. Commun.* **2018**, *9*.

(7) Paik, E. Y.; Zhang, L.; Burg, G. W.; Gogna, R.; Tutuc, E.; Deng, H. Interlayer Exciton Laser of Extended Spatial Coherence in Atomically Thin Heterostructures. *Nat. 2019 5767785* **2019**, *576*, 80–84.

(8) Jiang, Y.; Chen, S.; Zheng, W.; Zheng, B.; Pan, A. Interlayer Exciton Formation, Relaxation, and Transport in TMD van Der Waals Heterostructures. *Light Sci. Appl.* **2021**, *10*, 1–29.

(9) Lin, K.-Q. A Roadmap for Interlayer Excitons. *Light Sci. Appl.* **2021**, *10*.

(10) Palacios-Berraquero, C.; Kara, D. M.; Montblanch, A. R. P.; Barbone, M.; Latawiec, P.; Yoon, D.; Ott, A. K.; Loncar, M.; Ferrari, A. C.; Atatüre, M. Large-Scale Quantum-Emitter Arrays in Atomically Thin Semiconductors. *Nat. Commun.* **2017**, *8*, 1–6.

(11) Raja, A.; Chaves, A.; Yu, J.; Arefe, G.; Hill, H. M.; Rigosi, A. F.; Berkelbach, T. C.; Nagler, P.; Schüller, C.; Korn, T.; Nuckolls, C.; Hone, J.; Brus, L. E.; Heinz, T. F.; Reichman, D. R.; Chernikov, A. Coulomb Engineering of the Bandgap and Excitons in Two-Dimensional Materials. *Nat. Commun.* **2017**, *8*, 1–7.

(12) Wang, G.; Chernikov, A.; Glazov, M. M.; Heinz, T. F.; Marie, X.; Amand, T.; Urbaszek, B. Colloquium: Excitons in Atomically Thin Transition Metal Dichalcogenides. *Rev. Mod. Phys.* **2018**, *90*, 021001.

(13) Seyler, K. L.; Rivera, P.; Yu, H.; Wilson, N. P.; Ray, E. L.; Mandrus, D. G.; Yan, J.; Yao, W.; Xu, X. Signatures of Moiré-Trapped Valley Excitons in MoSe2/WSe2 Heterobilayers. *Nature* **2019**, *567*, 66–70.

(14) Sternbach, A. J.; Chae, S. H.; Latini, S.; Rikhter, A. A.; Shao, Y.; Li, B.; Rhodes, D.; Kim, B.; Schuck, P. J.; Xu, X.; Zhu, X. Y.; Averitt, R. D.; Hone, J.; Fogler, M. M.; Rubio, A.; Basov, D. N. Programmable Hyperbolic Polaritons in van Der Waals Semiconductors. *Science (80-. ).* **2021**, *371*, 617–620.

(15) Schneider, C.; Glazov, M. M.; Korn, T.; Höfling, S.; Urbaszek, B. Two-Dimensional Semiconductors in the Regime of Strong Light-Matter Coupling. *Nat. Commun. 2018 91* **2018**, *9*, 1–9.

(16) As'Ham, K.; Al-Ani, I.; Huang, L.; Miroshnichenko, A. E.; Hattori, H. T. Boosting Strong Coupling in a Hybrid WSe2Monolayer-Anapole-Plasmon System. *ACS Photonics* **2021**, *8*, 489–496.

(17) Al-Ani, I. A. M.; As'Ham, K.; Klochan, O.; Hattori, H. T.; Huang, L.; Miroshnichenko, A. E. Recent Advances on Strong Light-Matter Coupling in Atomically Thin TMDC Semiconductor Materials. *J. Opt.* **2022**, *24*, 053001.

(18) Moroz, A. Non-Radiative Decay of a Dipole Emitter Close to a Metallic Nanoparticle: Importance of Higher-Order Multipole Contributions. *Opt. Commun.* **2010**, *283*, 2277–2287.

(19) Rivera, P.; Fryett, T. K.; Chen, Y.; Liu, C. H.; Ray, E.; Hatami, F.; Yan, J.; Mandrus, D.; Yao, W.; Majumdar, A.; Xu, X. Coupling of Photonic Crystal Cavity and Interlayer Exciton in Heterobilayer of Transition Metal Dichalcogenides. *2D Mater.* **2019**, *7*, 015027.

(20) Förg, M.; Colombier, L.; Patel, R. K.; Lindlau, J.; Mohite, A. D.; Yamaguchi, H.; Glazov, M. M.; Hunger, D.; Högele, A. Cavity-Control of Interlayer Excitons in van Der Waals Heterostructures. *Nat. Commun.* **2019**, *10*.

(21) May, M. A.; Jiang, T.; Du, C.; Park, K. D.; Xu, X.; Belyanin, A.; Raschke, M. B. Nanocavity Clock Spectroscopy: Resolving Competing Exciton Dynamics in WSe2/MoSe2 Heterobilayers. *Nano Lett.* **2021**, *21*, 522–528.

(22) Rivera, P.; Schaibley, J. R.; Jones, A. M.; Ross, J. S.; Wu, S.; Aivazian, G.; Klement, P.; Seyler, K.; Clark, G.; Ghimire, N. J.; Yan, J.; Mandrus, D. G.; Yao, W.; Xu, X. Observation of Long-Lived Interlayer Excitons in Monolayer MoSe2–WSe2 Heterostructures. *Nat. Commun. 2015 61* **2015**, *6*, 1–6.

(23) Nagler, P.; Plechinger, G.; Ballottin, M. V.; Mitioglu, A.; Meier, S.; Paradiso, N.; Strunk, C.; Chernikov, A.; Christianen, P. C. M.; Schüller, C.; Korn, T. Interlayer Exciton Dynamics in a Dichalcogenide Monolayer Heterostructure. *2D Mater.* **2017**, *4*, 0–9.

(24) He, Z.; Han, Z.; Yuan, J.; Sinyukov, A. M.; Eleuch, H.; Niu, C.; Zhang, Z.; Lou, J.; Hu, J.; Voronine, D. V.; Scully, M. O. Quantum Plasmonic Control of Trions in a Picocavity with Monolayer WS2. *Sci. Adv.* **2019**, *5*.

(25) Ferrera, M.; Rahaman, M.; Sanders, S.; Pan, Y.; Milekhin, I.; Gemming, S.; Alabastri, A.; Bisio, F.; Canepa, M.; Zahn, D. R. T. Controlling Excitons in the Quantum Tunneling Regime in a Hybrid Plasmonic/2D Semiconductor Interface. *Appl. Phys. Rev.* **2022**, *9*, 031401.

(26) Savage, K. J.; Hawkeye, M. M.; Esteban, R.; Borisov, A. G.; Aizpurua, J.; Baumberg, J. J. Revealing the Quantum Regime in Tunnelling Plasmonics. *Nat. 2012 4917425* **2012**, *491*, 574–577.

(27) Rahaman, M.; Wagner, C.; Mukherjee, A.; Lopez-Rivera, A.; Gemming, S.; Zahn, D. R. T. Probing Interlayer Excitons in a Vertical van Der Waals P-n Junction Using a Scanning Probe Microscopy Technique. *J. Phys. Condens. Matter* **2019**, *31*.

(28) Park, K. D.; Jiang, T.; Clark, G.; Xu, X.; Raschke, M. B. Radiative Control of Dark Excitons at Room Temperature by Nano-Optical Antenna-Tip Purcell Effect. *Nat. Nanotechnol.* **2018**, *13*, 59–64.

(29) Rahaman, M.; Selyshchev, O.; Pan, Y.; Schwartz, R.; Milekhin, I.; Sharma, A.; Salvan, G.; Gemming, S.; Korn, T.; Zahn, D. R. T. Observation of Room-Temperature Dark Exciton Emission in Nanopatch-Decorated Monolayer WSe2 on Metal Substrate. *Adv. Opt. Mater.* **2021**, 2101801.

(30) Förster, T. Zwischenmolekulare Energiewanderung Und Fluoreszenz. *Ann. Phys.* **1948**, *437*, 55–75.

(31) Palummo, M.; Bernardi, M.; Grossman, J. C. Exciton Radiative Lifetimes in Two-Dimensional Transition Metal Dichalcogenides. *Nano Lett.* **2015**, *15*, 2794–2800.

(32) Suh, J. Y.; Kim, C. H.; Zhou, W.; Huntington, M. D.; Co, D. T.; Wasielewski, M. R.; Odom, T. W. Plasmonic Bowtie Nanolaser Arrays. *Nano Lett.* **2012**, *12*, 5769–5774.

(33) Wei, W.; Yan, X.; Zhang, X. Ultrahigh Purcell Factor in Low-Threshold Nanolaser Based on Asymmetric Hybrid Plasmonic Cavity. *Sci. Reports 2016 61* **2016**, *6*, 1–7.

(34) Park, K. D.; Raschke, M. B. Polarization Control with Plasmonic Antenna Tips: A Universal Approach to Optical Nanocrystallography and Vector-Field Imaging. *Nano Lett.* **2018**, *18*, 2912–2917.

(35) Behr, N.; Raschke, M. B. Optical Antenna Properties of Scanning Probe Tips: Plasmonic Light Scattering, Tip-Sample Coupling, and near-Field Enhancement. *J. Phys. Chem. C* **2008**, *112*, 3766–3773.

(36) Jang, A. R.; Hong, S.; Hyun, C.; Yoon, S. I.; Kim, G.; Jeong, H. Y.; Shin, T. J.; Park, S. O.; Wong, K.; Kwak, S. K.; Park, N.; Yu, K.; Choi, E.; Mishchenko, A.; Withers, F.; Novoselov, K. S.; Lim, H.; Shin, H. S. Wafer-Scale and Wrinkle-Free Epitaxial Growth of Single-Orientated Multilayer Hexagonal Boron Nitride on Sapphire. *Nano Lett.* **2016**, *16*, 3360–3366.

# Supplementary information

## Table of contents

## I. Sample preparation

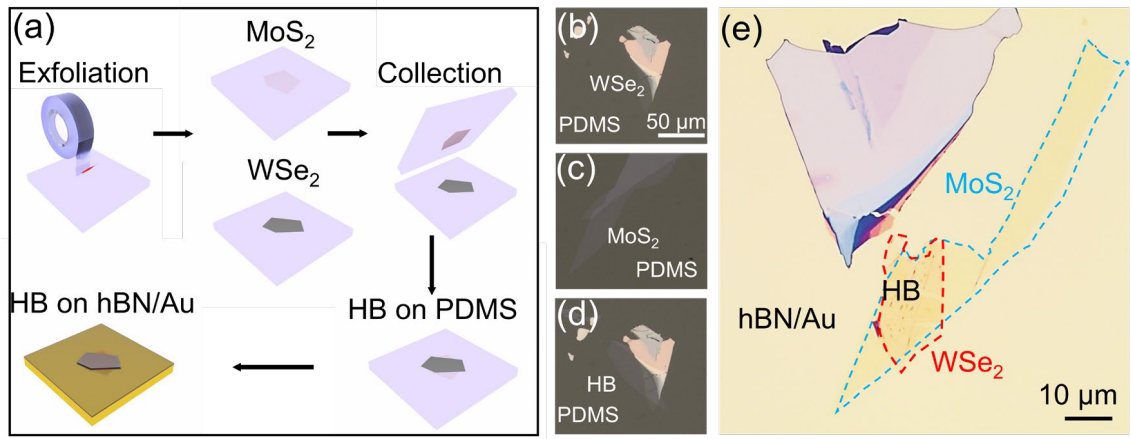

Fig. S1: (a) Schematic presentation of the HB preparation on hBN/Au substrate. We used a deterministic dry transfer method for the sample preparation. At first, both $MoS_2$ and $WSe_2$ monolayers were exfoliated on PDMS separately from the commercially available bulk crystals purchased from 2D semiconductors. After that, monolayer $MoS_2$ was picked up directly from the PDMS by $WSe_2$/PDMS. HBs prepared in this way preserve an ultraclean residuals-free interface which exhibits strong interlayer exciton formation as shown in Fig.1 in the main text. After picking up, HBs are then transferred onto hBN/Au substrate. (b) – (d) Optical microscope images of the $WSe_2$, $MoS_2$, and the picked-up HB on PDMS respectively. (e) Optical microscope image of the HB on hBN/Au substrate.

## II. Kelvin probe force microscopy characterization of HBs

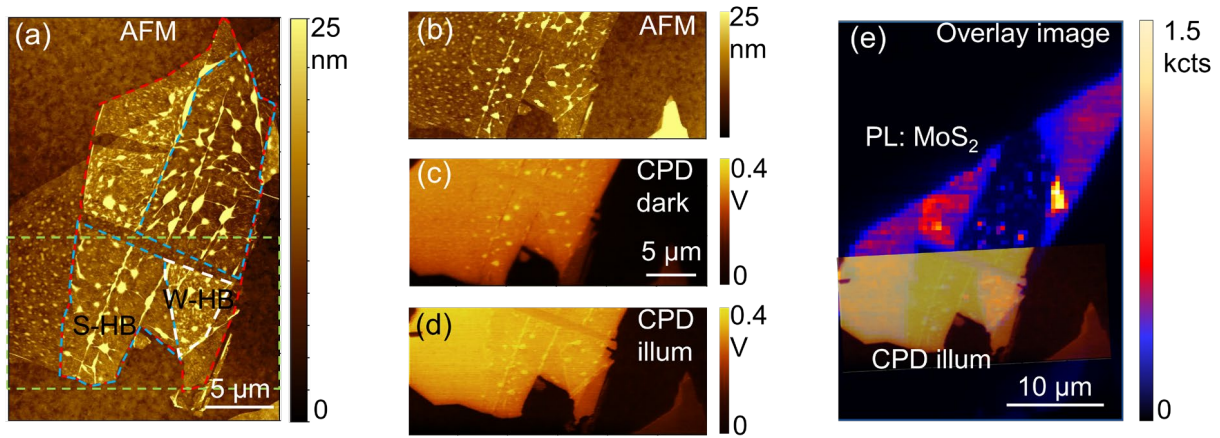

Fig. S2: (a) AFM image of the WSe$_2$/MoS$_2$ HB on hBN/Au substrate. Red dashed lines mark the area of WSe$_2$ flake, blue dashed lines highlight the clean HB area (strong coupling interface), and white dashed lines mark the unclean HB area (weak coupling interface) respectively. Here, WSe$_2$ monolayer is on top of MoS$_2$ monolayer. (b) – (d) AFM, contact potential difference (CPD) under dark and illumination with 633 nm laser of the HB area highlighted by the green dashed lines in (a) respectively. The laser power was 16 µW for the CPD measurement. Comparing the CPD images shown in (b) and (c), one can see that the clean interface area of the HB is clearly visible in the CPD image acquired under illumination with 633 nm. There is a CPD drop of ~120 meV for WSe$_2$ in the S-HB area, which clearly indicates the photogenerated charge transfer phenomena across the clean interface and the consequent drifting down of the fermi level of WSe$_2$ by ~120 meV. (e) Overlay image of CPD under illumination and the PL map of MoS$_2$ for comparison.

## III. Conductive AFM imaging of hBN/Au substrate

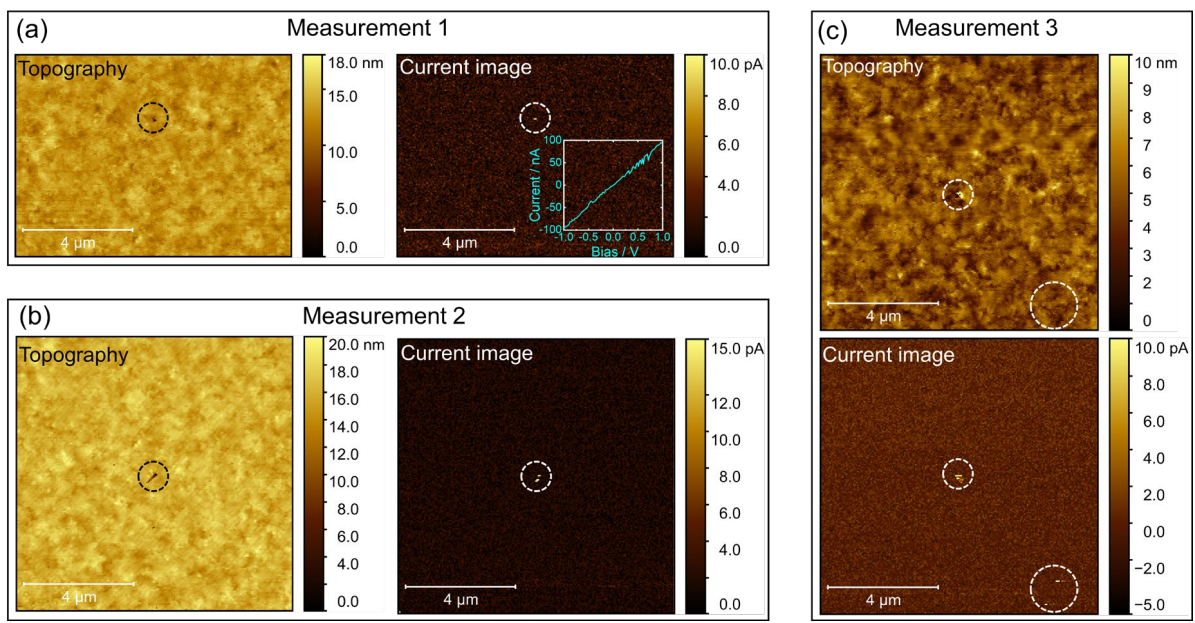

Fig. S3: (a) – (c) Three current maps together with corresponding AFM topography on hBN/Au substrate acquired using conductive AFM near the HB. A commercially available Au probe was used for the measurements. All the current images were taken at 0.2 V bias applied to the tip and the sample was grounded. An I – V sweep acquired on the high current point in (a) is presented in the inset of the current image in (a). We observed an ohmic-like I – V relation between the tip and the sample, which shows a direct conducting channel through the torn hBN film.

## IV. Tip-sample distance calculation

Our near-field measurements were performed by the Horiba NanoRaman platform. Fig. S4a shows one representative measured force *vs.* piezo displacement (*F vs. D*) curve during the tip-sample distance-dependent TEPL measurements using the setup. To convert the nominal force (expressed in a.u.) to actual force in nN, we first determined the deflection of the cantilever from the as-measured *F vs. D* curve via a linear fit to the repulsive regime. Interaction between the tip and the sample in the repulsive regime results in the elastic bending of the cantilever. Thus, we used Hook's law, $F = -k \cdot x$ with $k$ being the spring constant (3 N/m) and $x$ being the deflection of the cantilever to determine the actual force vs piezo displacement curve.

As shown in Fig. S4a, *F vs. D* curve has two interaction regimes: attractive (blue shed) and repulsive (red shed) regimes. In the repulsive regime, the actual tip-sample distance, $d$ can be calculated from the van der Waals force using the equation,

$$F_{rep}\ (nN) = \alpha d^{-13} \tag{1}$$

with $\alpha = 2.2 \times 10^{-7}\,\text{nN} \cdot \text{nm}^{-13}$ taken from the literature[1].

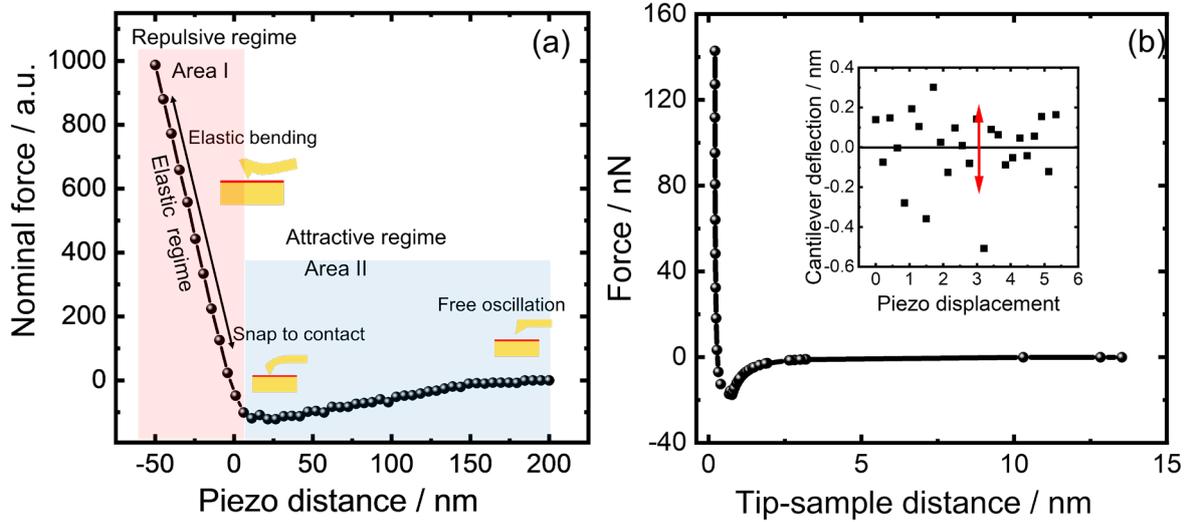

Fig. S4: (a) As measured nominal force (a.u.) *vs.* piezo displacement curve. There are two tip-sample interaction regimes highlighted by blue and red shades respectively in the curve. (b) The actual force *vs.* tip-sample distance (*F vs. d*) curve calculated from the *F vs D* curve is shown in (a). Inset is the standard deviation of the cantilever deflection before physical tip-sample contact.

Total tip-sample interaction force can be expressed in terms of attractive and repulsive forces as

$$F_{tot} = F_{att} + F_{rep} \tag{2}$$

Here, the attractive force can be written as the combination of[2]

$$F_{att} = -\frac{HR}{6d^2} + F_{cap} \qquad (3)$$

with $H$ is the Hamaker constant of Au ($2.5 \times 10^{-19}$ J) (ref.[3]), $R$ is the radius of the tip (25 nm), and $F_{cap}$ is the capillary force. To determine the actual tip-sample distance in the attractive regime we used the equation (3). Since both the Au tip and TMDCs are hydrophobic and the relative humidity in the experiment room was below 30 % during the experiment, we ignored $F_{cap}$ for the calculation. Capillary force scaled as $d/h$ with $h$ being the thickness of the water layer on the film[2]. For 2D materials including TMDCs water molecules tend to form droplets at high relative humidity (> 80 % RH) and show no sign of water structure ≤ 30 % RH[4]. Moreover, for a hydrophobic-hydrophobic system, capillary friction becomes insensitive below 40 % RH[5]. Therefore, ignoring the contribution from $F_{cap}$ in the above equation can still results in determination of tip-sample distance with high accuracy.

The uncertainty of the tip-sample distance determination in the repulsive regime using the equation (1) is mainly due to the fitting parameters which are smaller than 1 %. However, before a physical contact is made between the tip and the sample (in the attractive regime) an additional error for the fluctuation of the cantilever deflection also needs to be considered. In our case, this fluctuation was within a standard deviation of 0.17 nm as shown in the inset of Fig. S4b. Additionally, the piezo actuator has an inherent deviation of 0.1 nm. Therefore, the total measurement error in the attractive regime is ~0.20 nm.

## V. Tip-sample distance-dependent TEPL spectra of HB with/without junction current

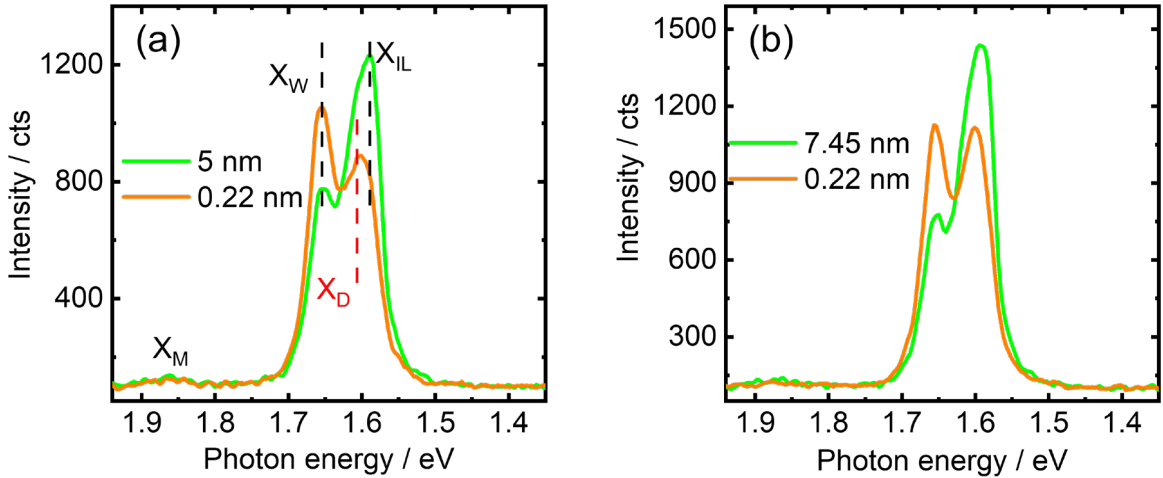

Fig. S5: Two representative tip-sample distance-dependent TEPL spectra (a) without junction current and (b) with junction current for the TEPL maps shown in Fig. 3 in the main text respectively. In addition to the intralayer $X_M$, $X_W$, and interlayer $X_{IL}$ we also observed intralayer dark exciton $X_D$ in WSe$_2$. $X_D$ becomes more apparent as the tip-sample distance decreases.

## VI. Rate equation model: radiative and nonradiative relaxation rate

In order to quantify the radiative and nonradiative relaxation as well as the Purcell factor as a function of tip-sample distance and junction current on/off, we developed a rate equation model adopted from the literature[6] to describe our tip-sample distance-dependent excitonic signals. The coupled rate equation can be written as

$$\frac{dN_W}{dt} = F - \left(\Gamma_W^{rad} + \Gamma_W^{nrad} + \Gamma_{CT}\right)N_W \quad (4)$$

Here, $N_W$ is the intralayer WSe$_2$ exciton ($X_W$) population, $\Gamma_W^{rad}$ and $\Gamma_W^{nrad}$ are the radiative and nonradiative relaxations, $\Gamma_{CT}$ is the interlayer charge transfer, and $F$ is the Purcell factor of the tip-sample cavity. $X_W$ population is proportional to the cavity enhancement as $F \propto |E/E_0|^2$ and scaled as $F \propto (R/z)^m$ with $R$ being the tip radius, $z$ being the distance between the tip and the Au substrate, and $m$ being the geometry-related field exponent factor. Both radiative and nonradiative decay is affected by the cavity field and is assumed to depend on the cavity volume by

$$\Gamma_W^{rad} \propto (z + z_0)^{-n} + \Gamma_W^{rad,0}$$

$$\Gamma_W^{nrad} \propto (R/(z + z_0))^l + \Gamma_W^{nrad,0} \quad (5)$$

Here, $z_0$ is the minimum tip-substrate distance, $\Gamma_W^{rad,0}$ and $\Gamma_W^{nrad,0}$ are the intrinsic unperturbed radiative and nonradiative decay into free space, and $n$, $l$ are the two scaling components. Using these assumptions coupled rate equation (4) can be solved in the steady state condition (which is true in our case since we used a low excitation laser power (17 μW)) to find the distance-dependent populations for $X_W$ in the following manner.

$$N_W = F / (\Gamma_W^{rad} + \Gamma_W^{nrad} + \Gamma_{CT})$$

$$P_W = \eta \Gamma_W^{rad} N_W \quad (6)$$

Here, $P_W$ is the PL intensity and $\eta \sim 1/Q$ with $Q$ being the quality factor of the cavity. The term $\eta$ cancels out $Q$ assuming that the cavity loss originates mainly from the radiative outcoupling into the far-field, in which radiative decay is scaled to $Q$ via $z$-dependent cavity volume[7]. Therefore, we can arrange Equation (6) into two parts: (i) in the large z range, where cavity field-induced radiative emission dominates and interlayer charge transfer is much faster than tip-induced nonradiative damping,

$$P_W = \eta F \Gamma_W^{rad} / \Gamma_{CT} \quad (7)$$

(ii) in the short z range ($z < 5$ nm), when tip-induced nonradiative damping overpower $\Gamma_{CT}$,

$$P_W = \eta F \Gamma_W^{rad} / \Gamma_W^{nrad} \quad (8)$$

Since in our case most important phenomena occur at a tip-sample distance $z < 5$ nm, we ignored the first scenario (equation (7)). Moreover, looking at the tip-sample distance-dependent PL evolution (see Fig. 4 in the main text), one can clearly see that we have two distinct regimes: (i) before the tip made contact with the sample and (ii) after the tip made contact with the sample. Therefore, we divided the PL evolution regime into two fitting zones and choose different scaling exponents for each zone respectively with all other parameters remaining the same. Table I and II presents all the fitting parameters used in the model.

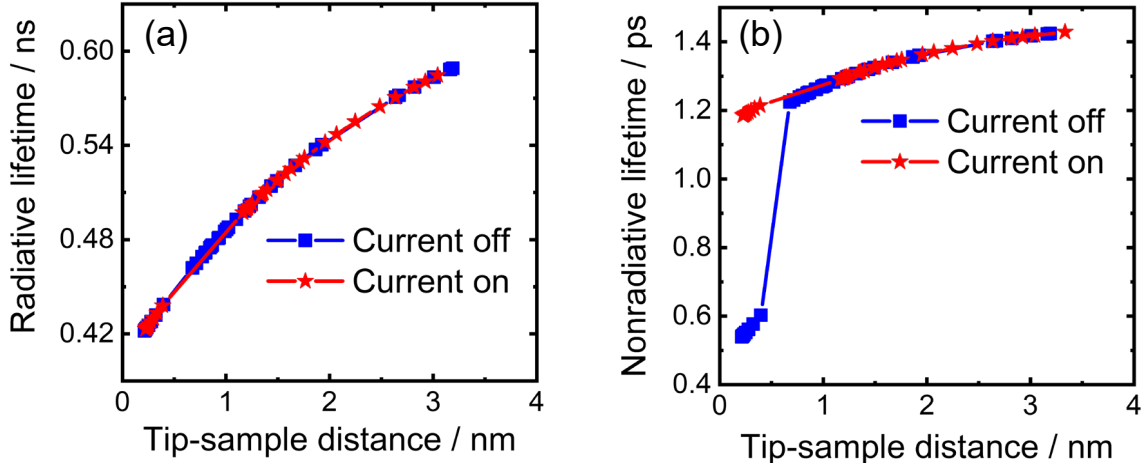

Fig. S6: (a) Radiative lifetime and (b) nonradiative lifetime of intralayer WSe$_2$ exciton (X$_W$) with/without junction current determined from the fitting of tip-sample distance dependent TEPL evolution shown in Fig. 4 in the main text using a rate equation.

Fig. S6 presents fitted radiative and nonradiative lifetimes of intralayer X$_W$ as a function of tip-sample distance. We observed that radiative lifetime both for current on/off decreases exponentially with the same scaling factor as the tip-sample gap shrinks. This is probably due to the fact that the in-plane polarized transition dipole of intralayer X$_W$ exciton is weakly coupled to the tip-sample cavity field polarized perpendicular to the HB basal plane[8,9]. However, tip-induced nonradiative damping for perpendicularly polarized exciton dipole exhibits strong $z$ dependency (follows $1/z^6$ dependency)[8]. Therefore, in the absence of a junction, current nonradiative damping becomes the dominant force. As a result, we observed a sharp decrease of interlayer exciton in the tip-sample sub-nm cavity with current off (Fig. 3f).

**Table I. Fitting parameters used in the rate equation model: current off.**

| Parameters | Before contact | After contact |
|---|---|---|
| m | 5.2 | 5.2 |
| n | 2.8 | 2.8 |
| l | 3.7 | 4.8 |
| $z_0$ | 4.4 nm | 4.4 nm |
| R | 25 nm | 25 nm |

**Table II. Fitting parameters used in the rate equation model: current on.**

| Parameters | Before contact | After contact |
|---|---|---|
| m | 5.4 | 4.9 |
| n | 2.8 | 2.8 |
| l | 3.7 | 3.7 |
| $z_0$ | 4.4 nm | 4.4 nm |
| R | 25 nm | 25 nm |

We assumed $\tau_W^{rad,0}$ and $\tau_W^{nrad,0}$ as 0.7 ns and 1.5 ps based on previous works[6,10]. Radiative and nonradiative relaxations were then calculated using $\tau = 2\hbar/\Gamma$.

**References**

(1) He, Z.; Han, Z.; Yuan, J.; Sinyukov, A. M.; Eleuch, H.; Niu, C.; Zhang, Z.; Lou, J.; Hu, J.; Voronine, D. V.; Scully, M. O. Quantum Plasmonic Control of Trions in a Picocavity with Monolayer WS2. *Sci. Adv.* **2019**, *5*.

(2) Zitzler, L.; Herminghaus, S.; Mugele, F. Capillary Forces in Tapping Mode Atomic Force Microscopy. *Phys. Rev. B* **2002**, *66*, 155436.

(3) Biggs, S.; Mulvaney, P. Measurement of the Forces between Gold Surfaces in Water by Atomic Force Microscopy. *J. Chem. Phys.* **1998**, *100*, 8501.

(4) Bampoulis, P.; Teernstra, V. J.; Lohse, D.; Zandvliet, H. J. W.; Poelsema, B. Hydrophobic Ice Confined between Graphene and MoS2. *J. Phys. Chem. C* **2016**, *120*, 27079–27084.

(5) Zhao, X.; Perry, S. S. The Role of Water in Modifying Friction within MoS2 Sliding Interfaces. *ACS Appl. Mater. Interfaces* **2010**, *2*, 1444–1448.

(6) May, M. A.; Jiang, T.; Du, C.; Park, K. D.; Xu, X.; Belyanin, A.; Raschke, M. B. Nanocavity Clock Spectroscopy: Resolving Competing Exciton Dynamics in WSe2/MoSe2Heterobilayers. *Nano Lett.* **2021**, *21*, 522–528.

(7) May, M. A.; Fialkow, D.; Wu, T.; Park, K.-D.; Leng, H.; Kropp, J. A.; Gougousi, T.; Lalanne, P.; Pelton, M.; Raschke, M. B.; May, M. A.; Park, K.; Raschke, M. B.; Fialkow, D.; Leng, H.; Kropp, J. A.; Gougousi, T.; Pelton, M.; Wu, T.; Lalanne, P. Nano-Cavity QED with Tunable Nano-Tip Interaction. *Adv. Quantum Technol.* **2020**, *3*, 1900087.

(8) Thomas, M.; Greffet, J. J.; Carminati, R.; Arias-Gonzalez, J. R. Single-Molecule Spontaneous Emission Close to Absorbing Nanostructures. *Appl. Phys. Lett.* **2004**, *85*, 3863.

(9) Carminati, R.; Greffet, J. J.; Henkel, C.; Vigoureux, J. M. Radiative and Non-Radiative Decay of a Single Molecule Close to a Metallic Nanoparticle. *Opt. Commun.* **2006**, *261*, 368–375.

(10) Palummo, M.; Bernardi, M.; Grossman, J. C. Exciton Radiative Lifetimes in Two-Dimensional Transition Metal Dichalcogenides. *Nano Lett.* **2015**, *15*, 2794–2800.